\documentclass[letterpaper,titlepage,11pt]{article}

\pdfoutput=1

\usepackage{amssymb,amsmath,amsfonts,jheppub}
\usepackage{mathrsfs,mathtools}
\usepackage[all]{hypcap}
\usepackage{amsmath,amssymb}
\usepackage{verbatim}
\usepackage{graphicx}
\usepackage{hyperref}
\usepackage{color}

\numberwithin{equation}{section}

\DeclareFontFamily{OT1}{rsfs}{}
\DeclareFontShape{OT1}{rsfs}{m}{n}{ <-7> rsfs5 <7-10> rsfs7 <10->rsfs10}{} 
\DeclareMathAlphabet{\mycal}{OT1}{rsfs}{m}{n}

\newcommand{\unity}{1\hspace{-0.243em}\text{l}}

\newcommand{\be}[1]{ \begin{equation}\label{#1} }
\newcommand{\ee}{\end{equation}}
\newcommand{\bea}[1]{\begin{eqnarray}\label{#1} }
\newcommand{\eea}{\end{eqnarray}}

\newcommand{\eq}[2]{\begin{equation} #1 \label{#2} \end{equation}}

\newcommand{\de}{\delta}

\DeclareMathOperator{\extdm}{d}
\newcommand{\extd}{\extdm \!}

\newcommand{\vp}{\varphi}
\newcommand{\cL}{\mathcal{L}}

\title{
Most general AdS$\boldsymbol{_3}$ boundary conditions
}

\subheader{TUW--16--16}

\author{Daniel Grumiller}
\author{and Max Riegler}

\affiliation{Institute for Theoretical Physics, TU Wien, Wiedner Hauptstr.~8-10/136, A-1040 Vienna, Austria}

\emailAdd{grumil@hep.itp.tuwien.ac.at}
\emailAdd{rieglerm@hep.itp.tuwien.ac.at}

\abstract{
We consider the most general asymptotically anti-de~Sitter boundary conditions in three-dimensional Einstein gravity with negative cosmological constant. The metric contains in total twelve independent functions, six of which are interpreted as chemical potentials (or non-normalizable fluctuations) and the other half as canonical boundary charges (or normalizable fluctuations). Their presence modifies the usual Fefferman--Graham expansion. The asymptotic symmetry algebra consists of two $\mathfrak{sl}(2)_k$ current algebras, the levels of which are given by $k=\ell/(4G_N)$, where $\ell$ is the AdS radius and $G_N$ the three-dimensional Newton constant.
}


\begin{document}

\maketitle

\section{Introduction}\label{se:1}

In this paper we reconsider Einstein gravity in three dimensions (3D) with negative cosmological constant, whose bulk action is given by
\eq{
I[g] = -\frac{1}{16\pi G_N}\,\int\extd^3x\sqrt{-g}\,\Big(R + \frac{2}{\ell^2}\Big)
}{eq:intro1}
where $\ell$ is the anti-de~Sitter (AdS) radius and $G_N$ the three-dimensional Newton constant. Our main goal is to find the loosest set of asymptotically AdS$_3$ boundary conditions (bc's).

Gravity in 3D is locally trivial \cite{Staruszkiewicz:1963zz, Deser:1984tn, Deser:1984dr}, so that most of the physics is determined by the bc's. The seminal example are Brown--Henneaux (BH) bc's, which involve essentially two arbitrary functions determining the physical state \cite{Brown:1986nw}. The bc preserving transformations split into trivial gauge transformations (those which leave the physical states intact) and asymptotic symmetries (those which transform one physical state into another). In the BH case the asymptotic symmetry algebra (ASA) consists of two copies of the Virasoro algebra with central charge $c=3\ell/(2G_N)$, i.e., the two-dimensional (2D) conformal algebra. 

Since then, the BH bc's were altered and generalized in numerous ways in 3D, e.g.~in the presence of scalar matter \cite{Henneaux:2002wm}, a gravitational Chern--Simons (CS) term \cite{Grumiller:2008es, Henneaux:2009pw, Skenderis:2009nt, Afshar:2011qw} or other higher derivative interactions \cite{Hohm:2010jc, Sinha:2010ai}. 
Even within locally AdS$_3$ Einstein gravity \eqref{eq:intro1} several alternatives to BH bc's were discovered that changed the ASA from the 2D conformal algebra to something else: a warped conformal algebra \cite{Compere:2013bya}, a centerless warped conformal algebra \cite{Donnay:2015abr}, a twisted warped conformal algebra \cite{Afshar:2015wjm} or the Heisenberg algebra \cite{Afshar:2016wfy}. As for BH, in these four alternatives the metric always has essentially two state-dependent functions that characterize the physical state. The same is true for the Korteweg--de~Vries bc's introduced very recently \cite{Perez:2016vqo}, where always two towers of canonical boundary charges emerge.

A few years ago Troessaert constructed more general bc's for 3D Einstein gravity that involve four state-dependent functions \cite{Troessaert:2013fma}. His ASA consists of two Virasoro and two $\mathfrak{u}(1)_k$ current algebras, which contains the other ASA's (conformal, warped or Heisenberg) as special cases (though not necessarily with the same central extensions). Also the canonical analysis of \cite{Troessaert:2015gra} led to four state-dependent functions. An independent set of bc's with four state-dependent functions was proposed in \cite{Avery:2013dja}, where the ASA contains a single $\mathfrak{sl}(2)_k$ current algebra (and a Virasoro algebra).

It is then fair to ask what is the upper limit, i.e., how many state-dependent functions can the metric contain at most, and what are the bc's that lead to this case? Moreover, what is the ASA for these most general bc's? In the present work we answer these questions and provide explicitly the loosest set of AdS$_3$ bc's for Einstein gravity.

Our conclusion that the ASA consists of two affine $\mathfrak{sl}(2)_k$ algebras is not surprising considering the usual relation between Wess--Zumino--Witten models and CS theories, which shows that the physical states in the spectrum are in a representation of the loop algebra [for $sl(2)$ this is the affine $\mathfrak{sl}(2)_k$], see e.g.~\cite{Elitzur:1989nr}. However, to the best of our knowledge this relation was never exploited for the purpose of establishing the loosest set of bc's in AdS$_3$ Einstein gravity. This is what we do in the present work.

This paper is organized as follows. In section \ref{se:2} we present our bc's in the CS formulation. In section \ref{se:3} we present the same bc's in the metric formulation. In section \ref{se:4} we conclude with some further checks, special cases, a potential loophole to full generality and comment on possible applications and generalizations to other dimensions.

\section{Boundary conditions in Chern--Simons formulation}\label{se:2}

Our goal is achieved more easily in the CS formulation, which is why we start with this formulation. In section \ref{se:2.1} we summarize salient features of the relation between CS theory and Einstein gravity \eqref{eq:intro1}, partly as a mini-review and partly to fix our notations and conventions. In section \ref{se:2.2} we present our new bc's. In section \ref{se:2.3} we determine the ASA.

\subsection{Notations and conventions}\label{se:2.1}

In the CS formulation of Einstein gravity, due to Ach\'ucarro, Townsend and Witten \cite{Achucarro:1987vz, Witten:1988hc}, the Einstein--Hilbert action \eqref{eq:intro1} is replaced with the difference of two CS actions
\begin{equation}
		I_{\textnormal{CS}}[A]-I_{\textnormal{CS}}[\bar{A}]
\label{eq:CS}
\end{equation}
where	
\begin{equation}
	I_{\textnormal{CS}}[A]=\frac{k}{4\pi}\int_{\mathcal{M}}\left<A\wedge\extd A+\tfrac{2}{3}A\wedge A\wedge A\right>
	\label{eq:cs}
\end{equation}
and the CS level is related to AdS radius and Newton's constant by
\eq{
k=\frac{\ell}{4G_N}\,.
}{eq:k}
To reduce clutter we set $\ell=1$ in this work. 

The connections $A, \bar A$ belong to the Lie algebra $sl(2,\mathbb{R})$. The equations of motion imply locally gauge flatness
\eq{
F=\extd A + A\wedge A=0=\extd\bar A + \bar A\wedge\bar A=\bar F
}{eq:eom}
which can be re-interpreted as the conditions of vanishing torsion, constant Ricci scalar and vanishing tracefree Ricci tensor, i.e., the 3D Einstein equations. Thus, the CS action \eqref{eq:CS} is classically equivalent to the Palatini action, which in turn is classically equivalent to the Einstein--Hilbert action \eqref{eq:intro1}.

Since AdS$_3$ has a cylinder as Penrose diagram we assume that our 3D manifold $\cal M$ topologically is a cylinder. When using coordinates explicitly, we are going to denote the radial coordinate of the cylinder by $\rho\in\mathbb{R}^+$, its angular coordinate by $\vp\sim\vp+2\pi$ and the coordinate along the cylinder by $t\in\mathbb{R}$. In Euclidean signature we assume that the manifold topologically is a filled torus with periodicities $(t,\,\vp)\sim(t,\,\vp+2\pi)\sim(t+2\pi,\,\vp)$.

We choose a standard basis to represent the three $sl(2,\mathbb{R})$ generators $L_{\pm1}$ and $L_0$
\begin{equation}
		[L_n,L_m]=(n-m)L_{n+m}
\end{equation}
and use the invariant bilinear form of $sl(2,\mathbb{R})$ in the fundamental representation.
\begin{equation}\label{eq:Sl2FundamentalKillingForm}
		\kappa_{ab}=\left<L_aL_b\right>=
			\begin{pmatrix}
				0&0&-1\\
				0&\frac{1}{2}&0\\
				-1&0&0
			\end{pmatrix}
	\end{equation}

In order to make contact with the metric formulation we quote one more result
\eq{
g_{\mu\nu}=\frac{1}{2}\left<(A_\mu-\bar{A}_\mu)(A_\nu-\bar{A}_\nu)\right>
}{eq:g}
which shows that the difference of the connections, $A-\bar A$, is essentially the dreibein (the sum $A+\bar A$ then determines the spin-connection). See \cite{Banados:1998gg,Carlip:2005zn,Campoleoni:2011tn,Riegler:2016PhD} for reviews and further references.

\subsection{Boundary conditions on the connection}\label{se:2.2}

Inspired by earlier constructions we partially gauge fix to radial gauge (see e.g.~\cite{Banados:1998gg})
\begin{equation}
A=b^{-1}\left[\extd + a(t,\varphi)\right]b\qquad\bar{A}=b\left[\extd+\bar{a}(t,\varphi)\right]b^{-1}
\label{eq:A}
\end{equation}
with a state-independent group element $b$ that we choose as\footnote{In this section the choice of $b$ is essentially irrelevant as neither the charges nor the ASA depend on it. A simpler, more standard, choice would be $b=\exp{(\rho L_0)}$. However, as we show below for the metric formulation the choice of $b$ is important and there are good reasons to pick \eqref{eq:b}, which we explain in section \ref{se:3}.}
\eq{
b=\exp{(L_{-1})}\,\exp{(\rho L_0)}\,.
}{eq:b}
In the ansatz \eqref{eq:A} we assume that all state-dependence is captured by the connection 1-forms $a$ and $\bar a$, which are independent from the radial coordinate.

We are now ready to present our bc's in the CS formulation.
	\begin{subequations}\label{eq:BoundaryConditions}
	\begin{align}
		a_\varphi&=-\frac{2\pi}{k}\left[\mathcal{L}^+(t,\varphi)L_{1}-2\mathcal{L}^0(t,\varphi)L_{0}+\mathcal{L}^-(t,\varphi)L_{-1}\right]\\
		\bar{a}_\varphi&=\frac{2\pi}{k}\left[\bar{\mathcal{L}}^+(t,\varphi)L_{1}-2\bar{\mathcal{L}}^0(t,\varphi)L_{0}+\bar{\mathcal{L}}^-(t,\varphi)L_{-1}\right]\\
		a_t&=\mu^+(t,\varphi) L_1+\mu^0(t,\varphi)L_0+\mu^-(t,\varphi)L_{-1}\\
		\bar{a}_t&=\bar{\mu}^+(t,\varphi) L_1+\bar{\mu}^0(t,\varphi)L_0+\bar{\mu}^-(t,\varphi)L_{-1}
	\end{align}
	\end{subequations}
Note that there are in total twelve independent functions. With hindsight, we call six of them ``charges'' (${\cal L}^a$, $\bar{\cal L}^a$) and six of them ``chemical potentials'' ($\mu^a$, $\bar\mu^a$), where $a=0,\pm 1$. The main difference between charges and chemical potentials is that only the former are allowed to vary:
	\begin{subequations}\label{eq:BoundaryConditions2}
	\begin{align}
		\delta a_\varphi&=-\frac{2\pi}{k}\left[\delta\mathcal{L}^+(t,\varphi)L_{1}-2\delta\mathcal{L}^0(t,\varphi)L_{0}+\delta\mathcal{L}^-(t,\varphi)L_{-1}\right]\\
		\delta\bar{a}_\varphi&=\frac{2\pi}{k}\left[\delta\bar{\mathcal{L}}^+(t,\varphi)L_{1}-2\delta\bar{\mathcal{L}}^0(t,\varphi)L_{0}+\delta\bar{\mathcal{L}}^-(t,\varphi)L_{-1}\right]\\
		\delta a_t&=\delta \bar{a}_t= 0
	\end{align}
	\end{subequations}
The allowed fluctuations $\delta A$ and $\delta\bar A$ follow from \eqref{eq:BoundaryConditions2} together with \eqref{eq:A} and \eqref{eq:b}.
	
For fixed chemical potentials the equations of motion \eqref{eq:eom} impose the following additional conditions on the charges $\mathcal{L}^a$ and $\bar{\mathcal{L}}^a$
	\begin{subequations}
	\label{eq:angelinajolie}
	\begin{align}
		\partial_t\mathcal{L}^0&=\mathcal{L}^-\mu^+-\mathcal{L}^+\mu^-+\frac{k}{4\pi}\partial_\varphi\mu^0 &
		\partial_t\mathcal{L}^\pm&=\pm\mathcal{L}^\pm\mu^0\pm2\mathcal{L}^0\mu^\pm-\frac{k}{2\pi}\partial_\varphi\mu^\pm\\
		\partial_t\bar{\mathcal{L}}^0&=\bar{\mathcal{L}}^-\bar{\mu}^+-\bar{\mathcal{L}}^+\bar{\mu}^--\frac{k}{4\pi}\partial_\varphi\bar{\mu}^0 &
		\partial_t\bar{\mathcal{L}}^\pm&=\pm\bar{\mathcal{L}}^\pm\bar{\mu}^0\pm2\bar{\mathcal{L}}^0\bar{\mu}^\pm+\frac{k}{2\pi}\partial_\varphi\bar{\mu}^\pm\,.
	\end{align}
	\end{subequations}

We shall demonstrate in the remainder of the paper that our bc's \eqref{eq:A}-\eqref{eq:BoundaryConditions2} pass all consistency tests. In particular they lead to finite and integrable canonical boundary charges, and allow for a well-defined variational principle.

\subsection{Asymptotic symmetry algebra}\label{se:2.3}

We consider now consequences of our bc's \eqref{eq:A}-\eqref{eq:BoundaryConditions2}. In particular, our main goal here is to derive the ASA through a canonical analysis. In the following we only focus on the $A$ sector, since the canonical analysis of the $\bar{A}$-sector works in complete analogy and yields the same results as the $A$-sector (upon decorating all functions with bars and replacing $k\rightarrow-k$ and $b\rightarrow b^{-1}$).

We start by considering all transformations
\eq{
\delta_\epsilon A = \extd\epsilon + [A,\,\epsilon] = {\cal O}(\delta A)
}{eq:epsilon}
that preserve our bc's \eqref{eq:A}-\eqref{eq:BoundaryConditions2}. To this end we split the gauge parameter $\epsilon$ into $sl(2,\mathbb{R})$-components.
	\begin{equation}
		\epsilon=b^{-1}\left[\epsilon_+(t,\varphi)L_1+\epsilon_0(t,\varphi)L_0+\epsilon_-(t,\varphi)L_{-1}\right]b
		\label{eq:epsilon2}
	\end{equation}
In fact, there is no restriction on the functions $\epsilon_a(t,\varphi)$, so that any transformation \eqref{eq:epsilon} with \eqref{eq:epsilon2} preserves our bc's \eqref{eq:A}-\eqref{eq:BoundaryConditions2}.

Thus we can already determine the infinitesimal changes of the state dependent functions $\mathcal{L}^\pm$ and $\mathcal{L}^0$ under bc-preserving transformations:
	\begin{subequations}\label{eq:InfinitesimalGaugeTrafos}
	\begin{align}
		\delta_\epsilon\mathcal{L}^\pm&=\pm\mathcal{L}^\pm\epsilon_0\pm2\mathcal{L}^0\epsilon_\pm-\frac{k}{2\pi}\partial_\varphi\epsilon_\pm\\
		\delta_\epsilon\mathcal{L}^0&=-\mathcal{L}^+\epsilon_-+\mathcal{L}^-\epsilon_++\frac{k}{4\pi}\partial_\varphi\epsilon_0
	\end{align}
	\end{subequations}
Since the chemical potentials $\mu^a$ are fixed we derive three constraints
	\begin{subequations}
	\label{eq:lalapetz}
	\begin{align}
		\delta_\epsilon\mu^\pm&=\pm\mu^\pm\epsilon_0\mp\mu^0\epsilon_\pm+\partial_t\epsilon_\pm=0\\
		\delta_\epsilon\mu^0&=2\mu^+\epsilon_--2\mu^-\epsilon_++\partial_t\epsilon_0=0
	\end{align}
	\end{subequations}
which fix the behavior of the gauge parameters $\epsilon_a$ under time evolution.

Using the background independent result for the variation of the canonical boundary charges \cite{Banados:1994tn,Banados:1998gg,Banados:1998ta,Carlip:2005zn}
\eq{
\delta{\cal Q}[\epsilon] = \frac{k}{2\pi}\,\oint\langle\epsilon\,\delta A\rangle
}{eq:Q}
obtains
	\begin{equation}
		\delta\mathcal{Q}[\epsilon]=\oint\extd\varphi\,\big(\delta\mathcal{L}^+\epsilon_-+\delta\mathcal{L}^0\epsilon_0+\delta\mathcal{L}^-\epsilon_+\big)
	\end{equation}
which can be functionally integrated to yield our final result for the canonical boundary charges
	\begin{equation}\label{eq:CanonicalBoundaryCharge}
		\mathcal{Q}[\epsilon]=\oint\extd\varphi\,\big(\mathcal{L}^+\epsilon_-+\mathcal{L}^0\epsilon_0+\mathcal{L}^-\epsilon_+\big)\,.
	\end{equation}

Having determined the infinitesimal transformations \eqref{eq:InfinitesimalGaugeTrafos} and the canonical boundary charge \eqref{eq:CanonicalBoundaryCharge} immediately yields the canonical realization of the (classical) ASA using standard methods \cite{Blagojevic}.
	\begin{equation}
		\{\mathcal{L}(t,\,\varphi)^a,\mathcal{L}(t,\,\bar{\varphi})^b\}=(a-b)\mathcal{L}^{a+b}(t,\,\varphi)\,\delta(\vp-\bar\vp)-\frac{k}{2\pi}\,\kappa_{ab}\,\partial_\varphi\delta(\varphi-\bar{\varphi})
	\end{equation}
Here $\{,\}$ denotes Dirac brackets. Choosing a suitable Fourier mode expansion and replacing Dirac brackets by commutators, $i\{,\}\rightarrow[,]$, yields
	\begin{equation}\label{eq:ASectorASA}
		[J^a_n,\,J^b_m]=(a-b)J_{n+m}^{a+b}-kn\,\kappa_{ab}\,\delta_{n+m,0}
	\end{equation}
where the bilinear form $\kappa_{ab}$ was defined in \eqref{eq:Sl2FundamentalKillingForm}. The algebra \eqref{eq:ASectorASA} is an affine $\mathfrak{sl}(2)_k$-algebra. 

Repeating the same analysis for the bar-sector our full ASA is then given by two copies of affine $\mathfrak{sl}(2)_k$-algebras.
\begin{subequations}
 \label{eq:ASA}
\begin{align}
 [J^a_n,\,J^b_m]&=(a-b)J_{n+m}^{a+b}-kn\,\kappa_{ab}\,\delta_{n+m,0}\\
 [\bar J^a_n,\,\bar J^b_m]&=(a-b)\bar J_{n+m}^{a+b}-kn\,\kappa_{ab}\,\delta_{n+m,0}
\end{align}
\end{subequations}
As mentioned in the introduction, the result \eqref{eq:ASA} may be expected on general grounds \cite{Elitzur:1989nr} and is compatible with the analysis of \cite{Banados:1994tn, Coussaert:1995zp}.

In the next section we translate the results above into the metric formulation, where a few subtleties arise that we shall expound upon.

\section{Boundary conditions in metric formulation}\label{se:3}

The bc's proposed in section \ref{se:2.2} straightforwardly translate into corresponding bc's on the dreibein or the metric, using the basic formulas reviewed in section \ref{se:2.1}. There is, however, a subtlety in the choice of the group element $b$ \eqref{eq:b} that we exhibit (and resolve) in section \ref{se:3.1}. In section \ref{se:3.2} we present our bc's in metric formulation. In section \ref{se:3.3} we discuss the asymptotic Killing vectors (AKVs) and their algebra.

\subsection{Generalized Fefferman--Graham gauge}\label{se:3.1}

As mentioned in section \ref{se:2} the choice of the group element $b$ appearing in the CS connection \eqref{eq:A} is irrelevant for the CS analysis (as long as $\delta b=0$). However, it becomes important for the metric interpretation. To see this imagine that we simply choose $b=\unity$. Then the CS analysis would be unchanged, but the metric would degenerate to a 2D metric, since $g_{\rho \mu}=0$ if $b=\unity$ [to check this simply plug \eqref{eq:A} with $b=\unity$ into \eqref{eq:g}]. This is why in the usual AdS$_3$ story one picks instead 
\eq{
\hat{b}=\exp{(\rho L_0)}\,. 
}{eq:bhat}

In order to appreciate our more complicated choice \eqref{eq:b} let us consider what happens for \eqref{eq:bhat} in our case.
The map \eqref{eq:g} yields
	\begin{subequations}\label{eq:MetricBoundarySoloCS}
	\begin{align}
		\hat g_{tt}&=\mu^+\bar{\mu}^-e^{2\rho}+\left[\tfrac{1}{4}\left(\mu^0-\bar{\mu}^0\right)^2-\mu^+\mu^--\bar{\mu}^+\bar{\mu}^-\right]+\mu^-\bar{\mu}^+e^{-2\rho} \displaybreak[1] \\
		\hat g_{t\varphi}&=\frac{\pi}{k}\left(\bar{\mathcal{L}}^-\mu^+-\mathcal{L}^+\bar{\mu}^-\right)e^{2\rho} +\frac{\pi}{k}\left[\mathcal{L}^-\mu^+-\bar{\mathcal{L}}^-\bar{\mu}^++\left(\mathcal{L}^0+\bar{\mathcal{L}}^0\right)(\mu^0-\bar{\mu}^0)+\mathcal{L}^+\mu^--\bar{\mathcal{L}}^+\bar{\mu}^-\right]\nonumber\\
		&\quad+\frac{\pi}{k}\left(\bar{\mathcal{L}}^+\mu^--\mathcal{L}^-\bar{\mu}^+\right)e^{-2\rho} \displaybreak[1] \\	
		\hat g_{t\rho}&=\tfrac{1}{2}\left(\mu^0-\bar{\mu}^0\right)  \displaybreak[1]\\
		\hat g_{\varphi\varphi}&=-\frac{4\pi^2}{k^2}\bar{\mathcal{L}}^-\mathcal{L}^+e^{2\rho}+\frac{4\pi^2}{k^2}\left[\left(\mathcal{L}^0+\bar{\mathcal{L}}^0\right)^2-\bar{\mathcal{L}}^-\bar{\mathcal{L}}^+-\mathcal{L}^-\mathcal{L}^+\right]-\frac{4\pi^2}{k^2}\bar{\mathcal{L}}^+\mathcal{L}^-e^{-2\rho}  \displaybreak[1]\\
		\hat g_{\varphi\rho}&=\frac{2\pi}{k}\left(\mathcal{L}^0+\bar{\mathcal{L}}^0\right)\\
		\hat g_{\rho\rho}&=1\,.
	\end{align}
	\end{subequations}
The crucial observation now is that the metric \eqref{eq:MetricBoundarySoloCS} only depends on four independent combinations of charges and four independent combinations of chemical potentials, which is two of each less than in the CS formulation. For instance, the information about the combination of chemical potentials $\mu^0+\bar\mu^0$ and charges $\cL^0-\bar\cL^0$ cannot be deduced from above.
Thus, for the choice of the group element \eqref{eq:bhat} some of the information contained in the CS connections \eqref{eq:BoundaryConditions} gets lost in the metric \eqref{eq:MetricBoundarySoloCS}. 

Whenever there is some loss of structure in the transition from CS to metric formulation it is conceivable that the group element $b$ was not chosen appropriately (see our example above where $b=\unity$). This motivated us to search for group elements different from \eqref{eq:bhat} that preserve all twelve functions in the metric. We have found several possible choices that do the job --- for instance, $b=\exp{(r L_1)}\exp{(-r L_{-1})}$ or $b=\exp{(L_{-1})}\exp{(\rho L_0)}$ or similar choices with $L_1\leftrightarrow L_{-1}$ or more complicated combinations thereof. Out of these choices we believe that \eqref{eq:b} leads to the simplest geometric interpretation of the asymptotically AdS$_3$ metrics, which is why we stick with it.

Our bc's \eqref{eq:A}-\eqref{eq:BoundaryConditions} then lead to the dreibein
\begin{multline}
 e^a_\mu\,L_a\,\extd x^\mu = \big[\big(\mu_1 e^\rho + \mu_2 e^{-\rho}\big)\extd t + \big(\cL_1 e^\rho+\cL_2 e^{-\rho}\big)\extd\vp\big]\,L_1  + \big[-\tfrac12\extd\rho+\big(\mu_4 e^\rho + \mu_5 + \mu_6 e^{-\rho}\big)\extd t \\
 + \big(\cL_4 e^\rho + \cL_5 + \cL_6 e^{-\rho}\big)\extd\vp\big]\,L_{-1} + \big[\extd\rho + \big(\mu_3-2\mu_2e^{-\rho}\big)\extd t +\big(\cL_3-2\cL_2e^{-\rho}\big)\extd\vp\big]\,L_0
\label{eq:e}
\end{multline}
where the $\mu_i$ ($\cL_i$) are related to the chemical potentials (charges) as follows.
\begin{subequations}\label{eq:AFVariablesFG}
	\begin{align}
		\mu_1&=\tfrac{1}{2}\mu^+ & \mu_2&=-\tfrac{1}{2}\bar{\mu}^+ & \mu_3&=\tfrac{1}{2}\left(2\mu^++\mu^0-\bar{\mu}^0\right) \\
		\mu_4&=-\tfrac{1}{2}\bar{\mu}^- & \mu_5 &=\tfrac{1}{2}\bar{\mu}^0 & \mu_6 &=\tfrac{1}{2}\left(\mu^+-\bar{\mu}^++\mu^0+\mu^-\right)\\
		\cL_1&=-\frac{\pi}{k}\mathcal{L}^+ & \cL_2&=-\frac{\pi}{k}\bar{\mathcal{L}}^+ & \cL_3 &=-\frac{2\pi}{k}\left(\mathcal{L}^+-\mathcal{L}^0-\bar{\mathcal{L}}^0\right) \\
		\cL_4 &=-\frac{\pi}{k}\bar{\mathcal{L}}^- & \cL_5 &=-\frac{2\pi}{k}\bar{\mathcal{L}}^0  & \cL_6& =-\frac{\pi}{k}\left(\mathcal{L}^++\bar{\mathcal{L}}^+-2\mathcal{L}^0+\mathcal{L}^-\right)
	\end{align}
\end{subequations}
From the equations above it is evident that all twelve functions contained in the connections $a, \bar a$ also appear in the dreibein. Since the map \eqref{eq:AFVariablesFG} between the metric variables $\mu_i, \cL_i$ and the CS variables $\mu^a, \bar\mu^a, \cL^a, \bar\cL^a$ is linear and invertible in the following we shall use either of these sets, depending on simplicity of the final result.

For reasons that will become apparent in the next subsection we call the choice \eqref{eq:b} ``generalized Fefferman--Graham gauge''. We are going to assume henceforth
\eq{
\mu_1\neq 0 \,. 
}{eq:assumptions}
This inequality guarantees that the dreibein \eqref{eq:e} has a leading $\extd t\,L_1$ component growing like $e^\rho$.

\subsection{Boundary conditions on the metric}\label{se:3.2}

Our gauge choice \eqref{eq:A} with \eqref{eq:b} yields the following generalized Fefferman--Graham expansion for the metric ($i,j=0,1$).
\begin{multline}
\extd s^2 = \extd\rho^2 + 2\,\Big(e^\rho N_i^{(0)} + N_i^{(1)} + e^{-\rho}N_i^{(2)} + {\cal O}(e^{-2\rho}) \Big)\, \extd\rho\extd x^i \\
+ \Big(e^{2\rho}\,g_{ij}^{(0)} + e^\rho\, g_{ij}^{(1)} + g_{ij}^{(2)} + {\cal O}(e^{-\rho})\Big)\, \extd x^i\extd x^j
\label{eq:FG}
\end{multline}
All expansion coefficients $g_{ij}^{(n)}$ depend on the boundary coordinates $x^i=(t,\vp)$, only.
Notably, the expansion coefficients $g_{ij}^{(1)}$ are non-zero. Moreover, the shift vector components $N_i^{(n)}$ cannot be removed by proper gauge transformations, in general. These are the key differences to the standard Fefferman--Graham expansion (see e.g.~\cite{Papadimitriou:2004ap} for a review). 

In terms of the generalized Fefferman--Graham expansion \eqref{eq:FG} our bc's on the metric are summarized in the next three sets of equations. The shift vector components $N_i^{(n)}$ are fixed as
\begin{subequations}
 \label{eq:bcN}
 \begin{align}
  N_t^{(0)} = g_{t\rho}^{(0)} &= \mu_1 \neq 0 &
  N_\vp^{(0)} = g_{\vp\rho}^{(0)} &= \cL_1 \\
  N_t^{(1)} = g_{t\rho}^{(1)} &= \mu_3 & 
  N_\vp^{(1)} = g_{\vp\rho}^{(1)} &= \cL_3 \\
  N_t^{(2)} = g_{t\rho}^{(2)} &= -\mu_2 &
  N_\vp^{(2)} = g_{\vp\rho}^{(2)} &= -\cL_2 \,.
 \end{align}
\end{subequations}
Above and below the functions $\mu_i(t,\vp)$ and $\cL_i(t,\vp)$ are expressed in terms of our original variables by virtue of the relations \eqref{eq:AFVariablesFG}.
The diagonal metric components $g_{ij}^{(n)}$ are fixed as
\begin{subequations}
 \label{eq:bcg}
 \begin{align}
  g_{tt}^{(0)} &= -4\mu_1\mu_4 & g_{\vp\vp}^{(0)} &= -4\cL_1\cL_4 \\
  g_{tt}^{(1)} &= -4\mu_1\mu_5 & g_{\vp\vp}^{(1)} &= -4\cL_1\cL_5 \\
  g_{tt}^{(2)} &= \mu_3^2-4\mu_1\mu_6-4\mu_2\mu_4 & g_{\vp\vp}^{(2)} &= \cL_3^2-4\cL_1\cL_6-4\cL_2\cL_4 \,.
 \end{align}
\end{subequations}
The off-diagonal metric components $g_{t\vp}$ are not independent from the expressions above, but determined by them algebraically through the Einstein equations.\footnote{%
In the special case $\cL_1=0$ we use \eqref{eq:gtphi} instead of the right equations \eqref{eq:bcg} to specify $\cL_4$, $\cL_5$ and $\cL_6$.
}
\begin{subequations}
 \label{eq:gtphi}
 \begin{align}
  g_{t\vp}^{(0)} &=  -2\mu_4\cL_1-2\mu_1\cL_4 \\
  g_{t\vp}^{(1)} &=  -2\mu_5\cL_1-2\mu_1\cL_5 \\
  g_{t\vp}^{(2)} &=  -2\mu_6\cL_1-2\mu_1\cL_6 - 2\mu_2\cL_4-2\mu_4\cL_2 + \mu_3\cL_3 \,.
 \end{align}
\end{subequations}
The Einstein equations $R_{\mu\nu}=-2g_{\mu\nu}$ impose additionally the on-shell constraints \eqref{eq:angelinajolie}. The allowed fluctuations of the metric are obtained from our assumptions $\delta\mu_i=0$ and $\delta\cL=$~arbitrary. They imply in particular
\eq{
\delta N_t = 0\qquad \delta g_{tt}=0\qquad \delta g_{t\vp}=\;\textrm{determined\;by\,}\delta N_\vp^{(0,1,2)}\;\textrm{and\;}\delta g_{\vp\vp}^{(0,1,2)}
}{eq:deltag}
where the three leading orders in $\delta N_\vp$ and $\delta g_{\vp\vp}$ are allowed to vary independently (they determine the variations of the six canonical boundary charges).

Note that the $t$- ($\vp$-) components of the shift vector and the $tt$- ($\vp\vp$-) components of the metric contain the whole information about the six chemical potentials (charges). This means that in general the shift vector cannot be eliminated by proper gauge transformations. 

Thus, the usual Fefferman--Graham gauge \cite{fefferman1985conformal} in general cannot be obtained from the generalized Fefferman--Graham gauge \eqref{eq:FG} by a proper gauge transformation, which implies that assuming Fefferman--Graham gauge as a starting point (like for instance in \cite{Skenderis:1999nb}) can only be achieved with loss of generality.

\subsection{Asymptotic Killing vectors}\label{se:3.3}

We determine now the AKVs, i.e., all vector fields $\xi^\mu$ with the property
\eq{
{\cal L}_\xi g_{\mu\nu} = {\cal O}(\delta g_{\mu\nu})
}{eq:Lie}
where ${\cal L}_\xi$ denotes the Lie-variation along $\xi$ and $\delta g_{\mu\nu}$ are the variations \eqref{eq:deltag} allowed by our bc's. With the ansatz
\eq{
\xi^\mu(t,\,\vp,\,\rho) = \xi^\mu_{(0)}(t,\,\vp) + e^{-\rho}\,\xi^\mu_{(1)}(t,\,\vp) + e^{-2\rho}\,\xi^\mu_{(2)}(t,\,\vp) + {\cal O}(e^{-3\rho})
}{eq:xiAnsatz}
we obtain from \eqref{eq:Lie} together with our bc's in section \ref{se:3.2} the results
\begin{subequations}
 \label{eq:xi}
 \begin{align}
  \xi^\vp_{(0)} &= \frac{k}{2\pi}\,\frac{\mu^+\bar{\epsilon}_--\bar{\mu}^-\epsilon_+}{\bar{\mathcal{L}}^-\mu^++\mathcal{L}^+\bar{\mu}^-} &
  \xi^t_{(0)} &= \frac{\mathcal{L}^+\bar{\epsilon}_-+\bar{\mathcal{L}}^-\epsilon_+}{\bar{\mathcal{L}}^-\mu^++\mathcal{L}^+\bar{\mu}^-} \\
  \xi^\vp_{(1)} &= \frac{k}{2\pi}\frac{\mu^+\bar{\lambda}^0}{\bar{\mathcal{L}}^-\mu^++\mathcal{L}^+\bar{\mu}^-} &
  \xi^t_{(1)} &= \frac{\mathcal{L}^+\bar{\lambda}^0}{\bar{\mathcal{L}}^-\mu^++\mathcal{L}^+\bar{\mu}^-} \\
  \xi^\vp_{(2)} &= -\frac{k}{2\pi}\frac{\mu^+\lambda^--\bar{\mu}^-\bar{\lambda}^+}{\bar{\mathcal{L}}^-\mu^++\mathcal{L}^+\bar{\mu}^-}&
  \xi^t_{(2)} &= -\frac{\mathcal{L}^+\lambda^-+\bar{\mathcal{L}}^-\bar{\lambda}^+}{\bar{\mathcal{L}}^-\mu^++\mathcal{L}^+\bar{\mu}^-}
 \end{align}
 \begin{align}
   \xi^\rho_{(0)} &= \tfrac{1}{2}\big[\epsilon_0-\bar{\epsilon}_0+2\epsilon_+-\left(\mu^0-\bar{\mu}^0-2\mu^+\right)\xi^t_{(0)}\big]-\frac{2\pi}{k}\left(\mathcal{L}^0+\bar{\mathcal{L}}^0-\mathcal{L}^+\right)\xi^\vp_{(0)} \\
   \xi^\rho_{(1)} &= \bar{\lambda}^+-\frac{2\left(\mathcal{L}^0+\bar{\mathcal{L}}^0\right)\mu^++\left(\mu^0-\bar{\mu}^0\right)\mathcal{L}^+}{2\left(\bar{\mathcal{L}}^-\mu^++\mathcal{L}^+\bar{\mu}^-\right)}\bar{\lambda}^0 \\
   \xi^\rho_{(2)} &=-\frac{2\pi}{k}\bar{\mathcal{L}}^+\xi^\vp_{(1)}-\bar{\mu}^+\xi^t_{(1)}+\bar{\lambda}^++\frac{\bar{\mathcal{L}}^-\left(\mu^0-\bar{\mu}^0\right)-2\bar{\mu}^-\left(\mathcal{L}^0+\bar{\mathcal{L}}^0\right)}{2\left(\bar{\mathcal{L}}^-\mu^++\mathcal{L}^+\bar{\mu}^-\right)}\bar{\lambda}^+\nonumber\\
    &\quad+\frac{2\mu^+\left(\mathcal{L}^0+\bar{\mathcal{L}}^0\right)+\mathcal{L}^+\left(\mu^0-\bar{\mu}^0\right)}{2\left(\bar{\mathcal{L}}^-\mu^++\mathcal{L}^+\bar{\mu}^-\right)}\lambda^-
 \end{align}
with
\begin{align}
		\bar{\lambda}^0&=-\frac{4\pi}{k}\mathcal{L}^0\xi^\varphi_{(0)}-\xi^\rho_{(0)}+\bar{\mu}^0\xi^t_{(0)}-\bar{\epsilon}_0\qquad\qquad \bar{\lambda}^+=-\frac{2\pi}{k}\bar{\mathcal{L}}^+\xi^\varphi_{(0)}-\bar{\mu}^+\xi^t_{(0)}+\bar{\epsilon}_+\\
		\lambda^-&=\frac{2\pi}{k}\left(2\bar{\mathcal{L}}^0\xi^\varphi_{(1)}+\left(\bar{\mathcal{L}}^++\mathcal{L}^-+\mathcal{L}^+-2\mathcal {L}^0\right)\xi^\varphi_{(0)}\right)+\xi^\rho_{(1)}-\bar{\mu}^0\xi^t_{(1)}\nonumber\\
		&-\left(\mu^+-\bar{\mu}^++\mu^0+\mu^-\right)\xi^t_{(0)}+\epsilon_0-\bar{\epsilon}_++\epsilon_-+\epsilon_+.
	\end{align}
\end{subequations}
Here $\epsilon_a(t,\vp)$ and $\bar\epsilon_a(t,\vp)$ with $a=0,\pm$ denote six arbitrary free functions. Note that the AKVs are state-dependent even to leading order. This state-dependence is crucial for obtaining the correct ASA.

The usual procedure when determining the ASA involves evaluating the Lie bracket between the AKVs
	\begin{equation}\label{eq:UsualLieBracket}
		[\xi_1,\xi_2]^\mu=\mathcal{L}_{\xi_1}\xi_2^\mu.
	\end{equation}
However, the expression \eqref{eq:UsualLieBracket} is only valid if the relevant pieces of the AKVs do not depend on state-dependent functions. 

Let us illustrate this with an example, where for concreteness we use Euclidean signature. Assume that the expansion coefficients $N^{(n)}_i$ and $g^{(n)}_{ij}$ are arbitrary functions of $t$ and $\vp$ that do not have the specific state-dependence as outlined in \eqref{eq:bcN} and \eqref{eq:bcg}. Then the AKVs that preserve these bc's are of the form
	\eq{
	\xi^t=f(t,\varphi)+\mathcal{O}(e^{-\rho})\qquad
	\xi^\varphi=g(t,\varphi)+\mathcal{O}(e^{-\rho})\qquad
	\xi^\rho=h(t,\varphi)+\mathcal{O}(e^{-\rho})
	}{eq:KillingVectors}
where the functions $f,\,g$ and $h$ are state-independent. Evaluating the Lie bracket \eqref{eq:UsualLieBracket} one finds that
	\begin{equation}
		[\xi(f_1,g_1,h_1),\xi(f_2,g_2,h_2)]^\mu=\xi^\mu(f_{[1,2]},g_{[1,2]},h_{[1,2]}),
	\end{equation}	
where $x_{[1,2]}=f_1\partial_tx_2+g_1\partial_\varphi x_2-f_2\partial_tx_1-g_2\partial_\varphi x_1$, with $x=f,g$ or $h$.
Introducing Fourier components as (we recall that in Euclidean signature $t\sim t+2\pi$)
	\begin{equation}
		F_{n|m}=\xi^\mu(e^{in\varphi+imt},0,0),\quad G_{n|m}=\xi^\mu(0,e^{in\varphi+imt},0),\quad H_{n|m}=\xi^\mu(0,0,e^{in\varphi+imt}),
	\end{equation}
one finds that these Fourier modes satisfy the following algebra:
	\begin{subequations}
	\begin{align}
		i[F_{n|p},F_{m|q}]&=(p-q)F_{n+m|p+q}\\
		i[F_{n|p},G_{m|q}]&=nF_{n+m|p+q}-qG_{n+m|p+q}\\
		i[F_{n|p},H_{m|q}]&=-qH_{n+m|p+q}\\
		i[G_{n|p},G_{m|q}]&=(n-m)G_{n+m|p+q}\\
		i[G_{n|p},H_{m|q}]&=-mH_{n+m|p+q}\\
		i[H_{n|p},H_{m|q}]&=0\,.
	\end{align}
	\end{subequations}
If one would try and use the same standard Lie bracket \eqref{eq:UsualLieBracket} for the state-dependent AKVs \eqref{eq:xi} then one would immediately encounter serious problems such as non-closure of the ASA. In order to fix these problems one has to modify \cite{Barnich:2010eb} (or ``adjust'' \cite{Compere:2015knw}) the Lie bracket \eqref{eq:UsualLieBracket} as follows   
	\begin{equation}\label{eq:ModLieBracket}
		[\xi_1,\xi_2]^\mu_M=\mathcal{L}_{\xi_1}\xi_2^\mu-\delta^g_{\xi_1}\xi_2^\mu+\delta^g_{\xi_2}\xi_1^\mu\,,
	\end{equation}
where $\delta^g_{\xi_1}\xi_2^\mu$ denotes the change induced in $\xi_2^\mu(g)$ due to the variation $\delta^g_{\xi_1}g_{\mu\nu}=\mathcal{L}_{\xi_1}g_{\mu\nu}$. Having the CS equivalent of the bc's \eqref{eq:bcN} and \eqref{eq:bcg} at hand one can immediately see that the changes of the state-dependent functions and the chemical potentials that appear in the AKVs \eqref{eq:xi} are essentially given by \eqref{eq:InfinitesimalGaugeTrafos} by the following general argument \cite{Witten:1988hc}.

One can explicitly verify that the AKVs \eqref{eq:xi} satisfy the relation
	\begin{equation}\label{eq:KillingEpsilonEpsilonbar}
		\epsilon-\bar\epsilon=2e_\mu\xi^\mu\,,
	\end{equation}
where $\epsilon$ and $\bar{\epsilon}$ are the gauge parameters that generated the bc preserving gauge transformations in the previous section and $e_\mu$ the dreibein \eqref{eq:e}. Using \eqref{eq:KillingEpsilonEpsilonbar} and $\left<[x,y],z\right>=\left<x,[y,z]\right>$, i.e., associativity of the invariant bilinear form one can readily show that, on-shell,
	\begin{equation}\label{eq:LieDerivativeMetricVariationRelation}
		\mathcal{L}_\xi g_{\mu\nu}=2\left(\left<\delta e_\mu,e_\nu\right>+\left<e_\mu,\delta e_\nu\right>\right)\,,
	\end{equation}
where we used the abbreviation $\delta e_\mu\equiv \frac{1}{2}\left(\delta_\epsilon A_\mu-\delta_{\bar{\epsilon}}\bar{A}_\mu\right)$. Thus the action of the Lie derivative on each component of the metric can be directly related with the infinitesimal gauge transformations of the gauge fields $A_\mu$ and $\bar{A}_\mu$ respectively which in the case at hand corresponds to \eqref{eq:InfinitesimalGaugeTrafos}.

A straightforward but tedious calculation using the modified Lie bracket \eqref{eq:ModLieBracket} yields 
	\begin{equation}
		[\xi(\{\epsilon_a^1,\bar{\epsilon}_a^1\}),\xi(\{\epsilon_a^2,\bar{\epsilon}_a^2\})]^\mu=\xi^\mu(\{\epsilon_a^{[1,2]},\bar{\epsilon}_a^{[1,2]}\})\,,
	\end{equation}
with
	\begin{subequations}
	\begin{align}
		\epsilon^{[1,2]}_\pm&=\pm\epsilon_0^1\epsilon_\pm^2\mp\epsilon_0^2\epsilon_\pm^1\qquad
		\epsilon^{[1,2]}_0=2\left(\epsilon_-^1\epsilon_+^2-\epsilon_-^2\epsilon_+^1\right)\\
		\bar{\epsilon}^{[1,2]}_\pm&=\pm\bar{\epsilon}_0^1\bar{\epsilon}_\pm^2\mp\bar{\epsilon}_0^2\bar{\epsilon}_\pm^1\qquad
		\bar{\epsilon}^{[1,2]}_0=2\left(\bar{\epsilon}_-^1\bar{\epsilon}_+^2-\bar{\epsilon}_-^2\bar{\epsilon}_+^1\right)
	\end{align}
	\end{subequations}
and $\{\epsilon_a,\bar{\epsilon}_a\}=(\epsilon_+,\epsilon_0,\epsilon_-,\bar{\epsilon}_+,\bar{\epsilon}_0,\bar{\epsilon}_-)$.
After introducing Fourier modes as
	\begin{subequations}
	\begin{align}
		J^+_{n|m}&=\xi^\mu(0_2,e^{in\varphi+imt},0_3) & J^0_{n|m}&=\xi^\mu(0,e^{in\varphi+imt},0_4) & J^-_{n|m}&=\xi^\mu(e^{in\varphi+imt},0_5)\\
		\bar{J}^+_{n|m}&=\xi^\mu(0_5,e^{in\varphi+imt}) & \bar{J}^0_{n|m}&=\xi^\mu(0_4,e^{in\varphi+imt},0) & \bar{J}^-_{n|m}&=\xi^\mu(0_3,e^{in\varphi+imt},0_2)
	\end{align}
	\end{subequations}
where $0_n$ denotes $n$ zeros (e.g.~$0_3=0,0,0$), one finds that these modes satisfy
	\begin{subequations}\label{eq:AKVAlgebraFromCS}
	\begin{align}
		[J^a_{n|p},J^b_{m|q}]&=(a-b)J^{a+b}_{n+m|p+q}\\
		[\bar{J}^a_{n|p},\bar{J}^b_{m|q}]&=(a-b)\bar{J}^{a+b}_{n+m|p+q}\\
		[J^a_{n|p},\bar{J}^b_{m|q}]&=0
	\end{align}
	\end{subequations}
which is essentially \eqref{eq:ASA} but without the central extensions, and with a double Fourier expansion with respect to $t$ and $\vp$. 
This shows the consistency of the algebra of AKVs with the canonical realization of the ASA \eqref{eq:ASA}. 

It is important to note that in order to make contact with the ASA found via the CS analysis one also has to fix the state dependence of the subleading parts of the AKVs. The reason for this is that while from a CS perspective all bc preserving gauge transformations are leading order contributions, some of these gauge transformations correspond to subleading contributions to the AKVs from the metric perspective.\footnote{Even though these contributions to the AKVs are subleading, their associated canonical charge is non-zero, as can be inferred from the CS formulation, and thus they do not generate proper gauge transformations.}

\section{Discussion}\label{se:4}

In this paper we have introduced the most general set of AdS$_3$ bc's possible in Einstein gravity with negative cosmological constant [subject to accessibility of radial gauge \eqref{eq:A}].
We conclude in this section with a discussion of our results.

In section \ref{se:4.1} we mention the checks that our bc's passed and discuss in addition conservation of the charges and consistency of the variational principle. In section \ref{se:4.2} we compare our bc's with previous special cases, namely BH, Comp\`ere--Song--Strominger, Heisenberg, Troessaert and Avery--Poojary--Suryanarayana bc's. In section \ref{se:loophole} we highlight a potential loophole to the generality of our bc's. In section \ref{se:4.3} we conclude with some comments on possible applications and generalizations to other dimensions.

\subsection{Checks}\label{se:4.1}

Our bc's \eqref{eq:A}-\eqref{eq:BoundaryConditions2} lead to canonical boundary charges \eqref{eq:CanonicalBoundaryCharge} that are non-trivial, finite, integrable and lead to an interesting ASA \eqref{eq:ASA} that matches with the algebra of AKVs \eqref{eq:AKVAlgebraFromCS}. 

We address now the conservation of the canonical boundary charges \eqref{eq:CanonicalBoundaryCharge} in time. The on-shell relations \eqref{eq:angelinajolie} together with the conditions on the time-derivatives of the gauge parameters \eqref{eq:lalapetz} yield the relation
\eq{
\partial_t{\mathcal Q}[\epsilon] = \frac{k}{2\pi}\,\oint\extd\vp\,\big(\epsilon_+\,\partial_\vp\mu^- + \epsilon_-\,\partial_\vp\mu^+ - \tfrac12\,\epsilon_0\,\partial_\vp\mu^0 \big)\,.
}{eq:conservation}
This implies conservation of the canonical boundary charges in time if the chemical potentials do not depend on the angular coordinate $\vp$. Otherwise, the canonical boundary charges change in time, but note that all state-dependent functions cancel in \eqref{eq:conservation}. 

Let us consider now the variational principle. It turns out that the bulk action \eqref{eq:cs} does not have a well-defined variational principle, i.e., the first variation of the bulk action does not vanish on-shell for some of the variations that preserve our bc's. Therefore, we need to add a suitable boundary term in order to restore a well-defined variational principle. This full bulk plus boundary action is given by
\eq{
\Gamma_{\textrm{\tiny CS}}=I_{\textrm{\tiny CS}} - \frac{k}{4\pi}\,\int_{\partial\mathcal M}\!\!\!\extd t\extd\vp\,\langle A_t A_\vp\rangle\,.
}{eq:vp}
A similar boundary term has to be added in the barred sector. Thus, our bc's lead to a well-defined variational principle 
\eq{
\delta\Gamma_{\textrm{\tiny CS}}\big|_{\textrm{\tiny EOM}} = - \frac{k}{2\pi}\,\int_{\partial\mathcal M}\!\!\!\extd t\extd\vp\,\langle a_\vp \,\delta a_t\rangle = 0
}{eq:checkvp}
provided the full action \eqref{eq:vp} is used. The expression on the right hand side of \eqref{eq:checkvp} has the familiar holographic form $\textrm{vev}\times\delta\,\textrm{source}$, with the six vev's ${\cal L}^a$ ($\bar{\cal L}^a$) contained in $a_\vp$ ($\bar a_\vp$) and the six sources $\mu^a$ ($\bar\mu^a$) contained in $a_t$ ($\bar a_t$). It could be an interesting exercise to establish a well-defined variational principle also in the metric formulation, which may require novel types of holographic counterterms different from the traditional ones \cite{Henningson:1998gx, Balasubramanian:1999re, Emparan:1999pm, deHaro:2000xn}.

\subsection{Previous special cases}\label{se:4.2}

Our bc's encompass all previous ones as special cases that require further restrictions on the state-dependent functions and/or chemical potentials. We mention here briefly how to obtain them in the context of our present work, starting with the recovery of bc's that have two state-dependent functions.

\subsubsection{Brown--Henneaux}

The bc's proposed in \cite{Brown:1986nw} are obtained from ours by further restricting
\eq{
\cL^+=-\bar\cL^-=-\frac{k}{2\pi}\qquad\cL^0=\bar\cL^0=0\qquad \cL^-,\,\bar\cL^+\;\textrm{arbitrary}\,.
}{eq:BH}

Again we restrict ourselves to one chiral sector, since the barred one is completely analogous. 
The on-shell conditions \eqref{eq:angelinajolie} then imply $\mu^0=-\partial_\vp\mu^+$, $\mu^-=-\frac{2\pi}{k}\mu^+\cL^- + \tfrac{1}{2}\partial_\vp^2\mu^+$ and (with $\cL:=\cL^-$, $\mu:=\mu^+$, dot denoting $\partial_t$ and prime denoting $\partial_\vp$)
\eq{
\dot\cL = \mu\cL^\prime + 2\mu^\prime\cL - \frac{k}{4\pi}\,\mu'''\,.  
}{eq:BH1}
Note that there is only one chemical potential in each chiral sector, $\mu^+$ and $\bar\mu^-$. In particular, the quantities $\mu^-$ and $\bar\mu^+$ are now state-dependent, which requires a different variational principle [namely, the boundary term subtracted in \eqref{eq:vp} must be set to zero].

The right hand side of \eqref{eq:BH1} shows the expected infinitesimal Schwarzian derivative with an anomalous term determined by $k$. As a consequence, the ASA is a Virasoro algebra with BH central charge $c=6k$ in each sector. Note that now there are only two canonical charges ($\cL, \bar\cL$) and two chemical potentials ($\mu, \bar\mu$).
Our functions $\mu_i, \cL_i$ can all be expressed in terms of $\mu, \bar\mu, \cL, \bar\cL$:
\begin{subequations}
 \label{eq:BH3}
 \begin{align}
   \mu_1&=\tfrac1 2\,\mu & \mu_2&=-\tfrac{2\pi}{k}\,\bar\mu\bar\cL-\tfrac{1}{2}\,\bar\mu'' & \mu_3&=- \tfrac12\,\big(\mu^\prime+\bar\mu^\prime\big) +\,\mu\\
   \mu_4&=-\tfrac{1}{2}\,\bar\mu & \mu_5&=\tfrac12\,\bar\mu^\prime & \mu_6&=\tfrac12\,\big(\mu-\tfrac{2\pi}{k}(\mu\cL+\bar\mu\bar\cL)-\mu'+\tfrac{1}{2}(\mu''-\bar\mu'')\big) \\
   \cL_1&=\tfrac12 & \cL_2&=-\tfrac{\pi}{k}\bar\cL & \cL_3&=1\\ 
   \cL_4&=-\tfrac12 & \cL_5&=0 & \cL_6&=-\tfrac{\pi}{k}\big(\bar\cL+\cL\big)+\tfrac12
 \end{align}
\end{subequations}
Note that the AKVs \eqref{eq:xi} become state-independent to leading order as a consequence of the restrictions \eqref{eq:BH}.

The main difference to the BH way of presenting their bc's is that we are using a group element $b$ \eqref{eq:b} that does not lead to a metric in Fefferman--Graham gauge. Explicitly, we obtain the metric \eqref{eq:FG} with the following expansion coefficients:

\begin{subequations}
\label{eq:BH2}
\begin{align}
N_t^{(0)}&=\mu_1 & N_t^{(1)}&=\mu_3 & N_t^{(2)}&=-\mu_2 \\
N_\vp^{(0)}&=\cL_1 & N_\vp^{(1)}&=\cL_3 & N_\vp^{(2)}&=-\cL_2 \\
g_{tt}^{(0)}&= -4\mu_1\mu_4 & g_{tt}^{(1)}&=-4\mu_1\mu_5 & g_{tt}^{(2)}&=\mu_3^2-4\mu_1\mu_6-4\mu_2\mu_4 \\
g_{\vp\vp}^{(0)}&= -4\cL_1\cL_4 & g_{\vp\vp}^{(1)}&=0 & g_{\vp\vp}^{(2)}&=\cL_3^2-4\cL_1\cL_6-4\cL_2\cL_4
\end{align}
\text{as well as}
\begin{align}
g_{t\vp}^{(0)} &=  -2\mu_4\cL_1-2\mu_1\cL_4\\
g_{t\vp}^{(1)} &=  -2\mu_5\cL_1\\
g_{t\vp}^{(2)} &=  -2\mu_6\cL_1-2\mu_1\cL_6 - 2\mu_2\cL_4-2\mu_4\cL_2 + \mu_3\cL_3\,.
\end{align}
\end{subequations}
The non-vanishing variations allowed by the BH bc's are given by
\eq{
\delta g_{tt}^{(2)},\,\delta g_{t\vp}^{(2)} = \;\textrm{arbitrary}\qquad \delta N_t^{(2)},\,\delta N_\vp^{(2)}=\;\textrm{determined}\,.
}{eq:BH4}
Explicitly, for constant chemical potentials $\mu=-\bar\mu=1$ the line-element reads
\begin{align}\label{eq:BH5}
\extd s^2 &= \extd\rho^2-\big(e^{2\rho}+\tfrac{2\pi}{k}(2\bar\cL-\cL)\big)\,\extd t^2 + \tfrac{2\pi}{k}\big(\cL+\bar\cL\big)\extd t\extd\vp + \big(e^\rho+2\big)\,\extd t\extd\rho\nonumber\\
&+ \big(e^{2\rho}+\tfrac{2\pi}{k}\cL\big)\extd\vp^2 + \big(e^\rho+2\big)\,\extd \vp\extd\rho + {\cal O}(e^{-\rho})
\end{align}
The metric \eqref{eq:BH5} obeys BH bc's in a slightly unusual coordinate system, which is a consequence of our choice \eqref{eq:b} for the group element.\footnote{%
The group element $b_{\textrm{\tiny gauge}}$ that generates the gauge transformation between our connection $A$ and the usual Brown--Henneaux connection $A_{\textrm{\tiny BH}}$ in highest weight gauge is given by $b_{\textrm{\tiny gauge}} = e^{-\rho L_0}e^{-L_{-1}}e^{\rho L_0}$, thus yielding
$A_{\textrm{\tiny BH}} = b_{\textrm{\tiny gauge}}^{-1} \,A\,b_{\textrm{\tiny gauge}} = e^{-\rho L_0}e^{L_{-1}}e^{\rho L_0} \, A\,  e^{-\rho L_0}e^{-L_{-1}}e^{\rho L_0}$.
The barred sector yields analogous results. In this gauge the metric takes the usual Fefferman--Graham form \cite{fefferman1985conformal}
$$\extd s^2_{\textrm{\tiny BH}} = \extd\rho^2 + e^{2\rho} (-\extd t^2 + \extd\vp^2) + M\,(\extd t^2 + \extd\vp^2) + 2J\,\extd t\extd\vp + {\cal O}(e^{-2\rho})$$ where in our conventions $M=\tfrac{2\pi}{k} (\cL-\bar\cL)$ and $J=\tfrac{2\pi}{k} (\cL+\bar\cL)$.
}

\subsubsection{Comp\`ere--Song--Strominger}

Another interesting set of bc's was proposed in \cite{Compere:2013bya}. They can be obtained from our bc's by relabelling $\varphi\rightarrow x^+$ and $t\rightarrow x^-$ as well as further restricting
\begin{equation}\label{eq:CSSStates}
    \cL^-=-\frac{\Delta}{k}\cL^+=-\frac{\Delta}{2\pi}\partial_+\bar{P}(x^+)\qquad\cL^0=\bar\cL^0=0\qquad \bar\cL^+=\frac{\bar{L}(x^+)}{2\pi}\qquad\bar\cL^-=-\frac{k}{2\pi}\,,
\end{equation}
where $\Delta$ is a fixed constant and $\bar{P}(x^+)$ and $\bar{L}(x^+)$ are arbitrary functions of their respective arguments. Note that the barred sector is equivalent to the BH case. In addition the chemical potentials are fixed as
\begin{equation}\label{eq:CSSChemPot}
    \mu^+=-\frac{k}{\Delta}\mu^-=1\qquad\mu^0=\bar\mu^a=0\,.
\end{equation}
Restricting our general bc's \eqref{eq:BoundaryConditions} in such a way also puts certain restrictions on the possible allowed gauge parameters \eqref{eq:epsilon2} as well as the infinitesimal variations \eqref{eq:InfinitesimalGaugeTrafos} of the remaining state dependent functions $\bar L$ and $\partial_+\bar P$. For the Comp\`ere--Song--Strominger bc's \eqref{eq:CSSStates} and \eqref{eq:CSSChemPot} this means that
\begin{subequations}
\begin{align}
    \epsilon_+&=\sigma-\left(\tfrac{1}{2}+\partial_+\bar P\right)\epsilon&\epsilon_0&=0&\epsilon_-&=\frac{\Delta}{k}\left(\left(\tfrac{1}{2}\partial_+\bar P\right)\epsilon-\sigma\right)\\
    \bar\epsilon_+&=\frac{\bar L}{k}\epsilon-\frac{\epsilon''}{2}&\bar\epsilon_0&=\epsilon'&\bar\epsilon_-&=-\epsilon
\end{align}
\end{subequations}
where $\epsilon\equiv\epsilon(x^+)$ and $\sigma\equiv\sigma(x^+)$. The general infinitesimal variations \eqref{eq:InfinitesimalGaugeTrafos} then simplify to
\begin{equation}
    \delta\left(\partial_+\bar P\right)=\epsilon'\left(\tfrac{1}{2}+\partial_+\bar P\right)+\epsilon\,\partial_+^2\bar P-\sigma'\qquad
    \delta\bar L=\epsilon\, \bar L'+2\epsilon'\,\bar L-\frac{k}{2}\epsilon'''\,.
\end{equation}
Then recombining the state dependent functions as
    \begin{equation}
        \mathfrak{L}:=\bar L-\Delta\left(\partial_+\bar P\right)^2\qquad\mathfrak{P}:=\Delta(1+2\partial_+\bar P)
    \end{equation}
one can easily verify that the Dirac brackets between the functions $\mathfrak{L}$ and $\mathfrak{P}$ are the ones of a $\mathfrak{u}(1)$ Kac-Moody-Virasoro algebra with a central charge $c=6k$ and a level\footnote{Here we use the same notion of $\mathfrak{u}(1)$ level as in \cite{Compere:2013bya} i.e.~$i\{\mathfrak{P}_n,\mathfrak{P}_m\}=\frac{\kappa}{2}n\,\delta_{n+m}$.} of the $\mathfrak{u}(1)$ current algebra $\kappa=-4\Delta$ thus showing that, indeed, the Comp\`ere--Song--Strominger bc's are a subset of our more general bc's.
The leading order contributions to the metric using our $b$ \eqref{eq:b} then read
\begin{align}
    \extd s^2&=\extd\rho^2+\frac{\Delta}{k}\big(\extd x^-\big)^2+\Bigg(e^{2\rho}\partial_+\bar P+\frac{\bar L(1+\partial_+ \bar P)+\Delta \big(\partial_+\bar P\big)^2}{k}\Bigg)\big(\extd x^+\big)^2\nonumber\\
    &-\Bigg(e^{2\rho}-\frac{\bar L-2\Delta\partial_+\bar P}{k}\Bigg)\extd x^+\extd x^-+\big(e^{2\rho}+2\big)\extd\rho\big(\extd x^+-\partial_+\bar P \extd x^-\big)+\ldots
\end{align}

In the remaining examples below we refrain from presenting the metric, since it always follows straightforwardly from our general results in section \ref{se:3}.

\subsubsection{Heisenberg}

The bc's proposed in \cite{Afshar:2016wfy} are obtained from ours by further restricting
\eq{
\cL^\pm=\bar\cL^\pm=0\qquad\cL^0,\,\bar\cL^0\;\textrm{arbitrary}
}{eq:H1}
and additionally assuming $\mu^\pm=\bar\mu^\pm=0$ (for simplicity). Again there are two charges ($\cL^0, \bar\cL^0$) and two chemical potentials ($\mu^0, \bar\mu^0$). The on-shell conditions \eqref{eq:angelinajolie} simplify to
\eq{
\dot\cL^0 = \frac{k}{4\pi}\,\mu^{0\,\prime}\qquad\dot{\bar\cL}^0 = -\frac{k}{4\pi}\,\bar\mu^{0\,\prime}
}{eq:H2}
and imply that the ASA consist of two $\mathfrak{u}(1)_k$ current algebras. However, when using the group element $b$ as given in \eqref{eq:b} we encounter the same type of problem that we discussed in section \ref{se:3.1}: the metric depends only on one combination of charges ($\cL^0+\bar\cL^0$) and one combination of chemical potentials ($\mu^0-\bar\mu^0$). The solution is the same as before: choose a more suitable $b$. Indeed, the choice used in \cite{Afshar:2016wfy} works and is given by $b=\exp{(\alpha L_1)}\exp{(\tfrac\rho2\,L_{-1})}$, where $\alpha$ is some state-independent constant.

This example highlights again the importance of choosing the group element $b$ in \eqref{eq:A} appropriately, i.e., in such a way that no functions are lost when translating the CS formulation into the metric formulation.

All cases above featured two state-dependent functions. Each of the remaining two examples below exhibits four state-dependent functions.

\subsubsection{Troessaert}

A more general set of bc's than the ones above was proposed in \cite{Troessaert:2013fma}. It encompasses the cases above in the sense that the corresponding ASA contains all the ASA's above as subalgebras (though not necessarily with the same central extensions). The Troessaert bc's allow for a fluctuating conformal factor in the leading order boundary metric as compared to BH bc's. One way to implement these boundary conditions\footnote{
We omit here a possible term linear in $t$ in the conformal factor used in \cite{Troessaert:2013fma}. For this reason our ASA will contain one generator less than the one in \cite{Troessaert:2013fma}.
} using our setup is by first switching from the canonical way of formulating boundary conditions to a holomorphic formulation. Using the CS formulation this can be done in a rather straightforward by by first relabelling $\varphi\rightarrow x^+$ and $t\rightarrow x^-$, i.e. switching to light cone coordinates as well as exchanging $\bar\cL^a\leftrightarrow\bar\mu^a$. The Troessaert bc's can then be realized by further restricting
\eq{
\cL^+=-\frac{k}{2\pi}\,e^{2\phi(x^+)}\qquad\bar\cL^- = \frac{k}{2\pi}\,e^{2\bar\phi(x^-)} \qquad\cL^0=\bar\cL^0=0\qquad \cL^-,\,\bar\cL^+\,\textrm{arbitrary}
}{eq:cedric1}
and setting all chemical potentials $\mu^a=\bar\mu^a=0$. The key (and only) differences to the BH bc's formulated in a holomorphic manner (see e.g.~\cite{Banados:1998gg}) are the factors $e^{2\phi(x^+)}$ and $e^{2\bar\phi(x^-)}$ that are allowed to vary arbitrarily. Like in the BH case we restrict ourselves to just one chiral sector of the theory.

Redefining our variables as (prime denotes differentiation with respect to $x^+$)
\eq{
{\cal L} = e^{2\phi}\,{\cal L}^-\qquad \epsilon = e^{-2\phi}\,\epsilon_+\qquad {\cal J} = -\frac{k}{2\pi}\,\phi^\prime
}{eq:cedric2}
yields the variations 
\begin{subequations}
 \label{eq:cedric4}
\begin{align}
\delta{\cal J} &= {\cal J}^\prime\epsilon + {\cal J}\epsilon^\prime - \frac{k}{4\pi}\,\epsilon'' - \frac{k}{4\pi}\,\epsilon_0^\prime \\
\delta{\cal L} &= {\cal L}^\prime\epsilon + 2{\cal L}\epsilon^\prime + \frac{k}{4\pi}\,\epsilon''_0 + {\cal J}\epsilon^\prime_0
\end{align}
\end{subequations}
and the canonical boundary charges
\eq{
{\cal Q} = \oint \extd\vp\, \big({\cal J}\,\epsilon_0 + {\cal L}\,\epsilon\big)\,.
}{eq:cedric3}
Introducing Fourier modes the variations \eqref{eq:cedric4} lead precisely to one chiral half of the ASA presented in Eq.~(6.8) of \cite{Troessaert:2013fma} (modulo the operator $Q$), with the same values for the anomalous terms. (The other chiral half of the ASA follows from our barred sector.)
\begin{subequations}
\label{eq:cedric5}
\begin{align}
[L_n,\,L_m] &= (n-m)\,L_{n+m}\\
[L_n,\,J_m] &= -m\,J_{n+m} + i\,\frac k2\,n^2\,\delta_{n+m,\,0} \label{eq:cedric5a}\\
[J_n,\,J_m] &= -\frac k2\,n\,\delta_{n+m,\,0}
\end{align}
\end{subequations}
As usual, the anomalous term in \eqref{eq:cedric5a} can be removed by a twisted Sugawara shift, see for instance \cite{Troessaert:2013fma} or \cite{Afshar:2015wjm}, where the algebra \eqref{eq:cedric5} arises as ASA in Rindleresque AdS$_3$ holography.

\subsubsection{Avery--Poojary--Suryanarayana}

The bc's proposed in \cite{Avery:2013dja} by Avery--Poojary--Suryanarayana lead to another interesting set of asymptotic symmetries in the form of a semidirect sum of a Virasoro algebra with central charge $c=6k$ and an $\mathfrak{sl}(2)_k$ algebra. This ASA can be obtained from our bc's by restricting either one of the two  sets of state-dependent functions and chemical potentials to the BH case \eqref{eq:BH} and no further constraints on the other set. Subjecting the unbarred sector to BH bc's yields
\eq{
\cL^+=-\frac{k}{2\pi}\qquad\cL^0=0\qquad \cL^-\,\textrm{arbitrary}\,.
}{eq:APS}
Renaming $\bar\cL^a\rightarrow\mathcal{T}^a$ we perform the Sugawara shift
\begin{equation}
    \cL:=\cL^-+\frac{2\pi}{k}\big(\mathcal{T}^0\mathcal{T}^0-\mathcal{T}^+\mathcal{T}^-\big)\,.
\end{equation}
Fourier expanding $\cL$ and $\mathcal{T}^a$ eventually establishes the ASA
\begin{subequations}
\begin{align}
    [L_n,\,L_m] &= (n-m)\,L_{n+m}+\frac{c}{12}\,n(n^2-1)\,\delta_{n+m,\,0}\\
    [L_n,\,T^a_m] &= -m\,T_{n+m}\\
    [T^a_n,\,T^b_m] &= (a-b)T^{a+b}_{n+m}-k\,n\,\kappa_{ab}\,\delta_{n+m,\,0}
\end{align}
\end{subequations}
with $c=6k$, which is exactly the ASA first presented in \cite{Avery:2013dja}.

This concludes our reproduction of previous bc's from our more general set of bc's and shows that indeed all previous constructions are contained as special cases.

\subsection{Loophole to generality}\label{se:loophole}

A potential loophole to full generality of our bc's is that one could allow gauges different from radial gauge or allow state-dependence of the group element $b$ in \eqref{eq:A}. We have no proof as of yet that there is no loss of generality in assuming \eqref{eq:A}, \eqref{eq:b}. However, since a counting of integration functions in the equations of motion \eqref{eq:eom} yields that there cannot be more than six state-dependent functions in $sl(2,\,\mathbb{R})\oplus sl(2,\,\mathbb{R})$ CS theory, and our bc's lead to six state-dependent functions we think that this is a very strong indication that there is no loss of generality. It would be nice to close this loophole by a proof.

In the following we sketch a possible proof.\footnote{
We thank Wout Merbis for discussions regarding state-dependence of $b$.
}
(See also the discussion in \cite{Banados:1994tn}.) If we assume the split \eqref{eq:A} with $a=a_t(t,\,\vp)\extd t+a_\vp(t,\,\vp)\extd\vp$ and $b=b(\rho,\,t)$, where $\delta b\neq 0$ in general, then the canonical boundary charges \eqref{eq:Q} acquire a $b$-dependent piece containing $(\delta b)b^{-1}$. Now either of these possibilities must arise as the boundary is approached: 1.~the new term diverges, 2.~the new term vanishes, 3.~the new term is finite. In case 1.~the boundary conditions are unphysical since the associated charges are infinite, so we can disregard it. In case 2.~the new term vanishes and therefore variations of $b$ can be removed by small gauge transformations. In case 3.~one can plausibly redefine $a$ to absorb all state-dependence of $b$, thereby recovering $\delta b=0$.

\subsection{Holographic interpretation}
Starting with the ASA \eqref{eq:ASA} following from our bc's  \eqref{eq:A}-\eqref{eq:BoundaryConditions2}, it is very suggestive that the corresponding dual field theory is a non-chiral $\mathfrak{sl}(2)_k$ WZW Model (see e.g.~\cite{Elitzur:1989nr}) since the physical states fall into representations of two affine $\mathfrak{sl}(2)_k$. The chemical potentials $\mu^a$ and $\bar\mu^a$ are then interpreted as sources that couple to the operators given by the left and right chiral $\mathfrak{sl}(2)_k$ currents $J^a$ and $\bar J^a$ whose vev's are given by the functions $\cL^a$ and $\bar\cL^a$ in \eqref{eq:BoundaryConditions}. 

It is illuminating to compare this holographic interpretation of our bc's with the holographic interpretation of the BH bc's \eqref{eq:BH}. In the BH case the (classical) physical state space is characterized by a holomorphic and an antiholomorphic function [$\cL(x^+)$ and $\bar\cL(x^-)$ in \eqref{eq:BH}], which appear in the canonical boundary charges. In the holographic interpretation these are the vev's of the corresponding operators. For our general bc's the situtaion is conceptually the same, except that instead of having 1+1 functions characterizing the canonical boundary charges we have 3+3 functions, $\cL^a$ and $\bar\cL^a$. 

Since the CS-formulation that we employed is classically equivalent to the metric formulation, everything we did in the former could be translated into the latter. We gave some examples, but clearly there are more aspects that could be translated, like the variational principle or holographic renormalization.\footnote{%
To give one more example, we identify here normalizable and non-normalizable linearized fluctuations around some background. That is, we split all 12 functions in the metric \eqref{eq:FG}-\eqref{eq:gtphi} into background and fluctuations, $\mu_i=\bar\mu_i+\de\mu_i$, $\cL_i=\bar\cL_i+\de\cL_i$, where barred quantities refer to the background; a simple choice is $\bar\mu_1=\bar\mu_4=\bar\cL_1=-\bar\cL_4=\tfrac12$ and all other $\bar\mu_i$ and $\bar\cL_i$ vanish, leading to the (asymptotically AdS$_3$) background line-element $\extd\bar s^2=\extd\rho^2+e^{2\rho}(-\extd t^2+\extd\vp^2)+e^\rho\extd\rho\,(\extd t +\extd\vp)$. Non-normalizable fluctuations (those, which violate our boundary conditions of fixed $\mu_i$) by definition have in general $\de\mu_i\neq 0$ (and, for simplicity, we may choose $\de\cL_i=0$ by adding suitable normalizable modes to the non-normalizable ones), while normalizable fluctuations have $\de\mu_i=0$ and, in general, $\de\cL_i\neq 0$. In the metric formulation there is then a simple way to discriminate normalizable from non-normalizable fluctuations $\delta g_{\mu\nu}$: if the linearized fluctuation $\delta g_{\mu\nu}$ is chosen such that it maintains the generalized Fefferman--Graham gauge \eqref{eq:FG} then in this gauge normalizability is the condition $\delta g_{tt}=\delta g_{t\rho}=0$. Note that all six charges $\cL_i$ enter in the normalizable fluctuations through the leading, subleading and sub-subleading components of $\delta g_{\vp\rho}$ and $\delta g_{\vp\vp}$ (and, redundantly, also of $\delta g_{t\vp}$).
} Generalizations to wormhole-like spacetimes along the lines of \cite{Skenderis:2009ju} could also be of interest, as well as the holographic calculation of $n$-point functions, along the lines of \cite{Bagchi:2015wna}.


\subsection{Towards applications and generalizations to other dimensions}\label{se:4.3}

The most general AdS$_3$ bc's \eqref{eq:A}-\eqref{eq:BoundaryConditions2} lead to two $\mathfrak{sl}(2)_k$ current algebras as ASA. We have seen above that all known special cases lead to ASA's that are certain subalgebras of ours. It could be interesting to classify all such subalgebras in order to get a full classification of all consistent AdS$_3$ bc's. It is possible that new sets of AdS$_3$ bc's can be discovered in this way that are different from the ones reviewed in section \ref{se:4.2}. 

Our bc's allow black holes as part of the spectrum, for instance BTZ black holes \cite{Banados:1992wn,Banados:1992gq}. It is then an interesting question whether the symmetries of our ASA \eqref{eq:ASA} allow again a Cardy-type of microstate counting of the entropy of these black holes. We leave this for future work.

It could be interesting to generalize our bc's to other cases in three dimensions, like flat space or asymptotically de Sitter, and to higher spin theories, either in an AdS$_3$ context \cite{Henneaux:2010xg, Campoleoni:2010zq} or for non-AdS$_3$ approaches \cite{Gary:2012ms, Afshar:2012nk} like flat space higher spin gravity \cite{Afshar:2013vka, Gonzalez:2013oaa}. Moreover, it could be rewarding to consider generalizations to other dimensions. For instance, it is conceivable that the AdS$_2$ bc's proposed in \cite{Grumiller:2013swa, Grumiller:2015vaa} can be generalized along the lines of our present work and may lead to a single copy of an $\mathfrak{sl}(2)_k$ current algebra as ASA. (For a relation between AdS$_3$ and AdS$_2$ gravity see \cite{Apolo:2014tua}.) Generalizations to higher dimensions, in particular four and five, would also be of interest. For this purpose, the presentation of our results in the metric formulation together with the generalized Fefferman--Graham-type of expansion \eqref{eq:FG} should be useful.

\acknowledgments

DG is grateful to C\'edric Troessaert for discussions in Valdivia in February 2016 and for sharing his expectation that the most general bc's in 3D could involve six state-dependent functions.
We also thank Max Ba\~nados, Glenn Barnich, Geoffrey Comp\`ere, St{\'e}phane Detournay, Wout Merbis, Blaza Oblak, Alfredo Perez, Massimo Porrati, Stefan Prohazka, Jakob Salzer, Friedrich Sch\"oller, Kostas Skenderis, David Tempo, Ricardo Troncoso, Shahin Sheikh-Jabbari and Dima Vassilevich for discussions.

DG acknowledges the hospitality of CECS Validivia, the Abdus Salam ICTP Trieste, the CTP at MIT, the Simons Center in Stony Brook, Brussels University and the International Solvay Institute, the Munich Institute for Astro- and Particle Physics (MIAPP), the IPM Teheran, the IMBM, Bo\u{g}azi\c{c}i University and Feza G\"ursey Center in Istanbul, the IIP in Natal, and the ABC Federal University in S\~ao Paulo during the course of this work.

MR acknowledges the hospitality of the Yukawa Institute for Theoretical Physics at the University of Kyoto during the course of this work.

This work was supported by the Austrian Science Fund (FWF), projects P~27182-N27 and P~28751-N27. DG was additionally supported by the program Science without Borders, project CNPq-401180/2014-0. MR was additionally supported by a DOC fellowship of the Austrian Academy of Sciences and the Doktoratskolleg ``Particles \& Interactions'' (FWF project DKW~1252-N27). 

\addcontentsline{toc}{section}{References}
\bibliographystyle{fullsort}
\bibliography{review}

\end{document}